\documentclass{emulateapj}

\slugcomment{Accepted for publication in ApJ.}
\shorttitle{Galaxy Classification}
\shortauthors{Bailin and Harris}

\defcitealias{shao-etal07}{S07}

\newcommand{\gr}{{\ensuremath{(g-r)}}}
\newcommand{\grfo}{{\ensuremath{(g-r)^{\mathrm{F}}}}}
\newcommand{\absmag}{{\ensuremath{M_r}}}
\newcommand{\absmagfo}{{\ensuremath{M_r^{\mathrm{F}}}}}
\newcommand{\cpetro}{{\ensuremath{C_{\mathrm{Petro}}}}}
\newcommand{\cnorm}{{\ensuremath{C_{\mathrm{norm}}}}}
\newcommand{\cmd}{{\ensuremath{\mathrm{CMD}}}}
\newcommand{\cmdfo}{{\ensuremath{\mathrm{CMD}^{\mathrm{F}}}}}

\begin{document}

\title{Inclination-Independent Galaxy Classification}

\author{Jeremy Bailin\altaffilmark{1} \and William E. Harris}
\affil{Department of Physics \& Astronomy, McMaster University,
  1280 Main Street West, Hamilton, ON, L8S 4M1, Canada}
\altaffiltext{1}{bailinj@mcmaster.ca}

\begin{abstract}
We present a new method to classify galaxies from large surveys
like the Sloan Digital Sky Survey using
inclination-corrected concentration, inclination-corrected location
on the color-magnitude diagram, and apparent axis ratio.
Explicitly accounting for inclination
tightens the distribution of each of these parameters and enables
simple boundaries to be drawn that delineate three different galaxy
populations: Early-type galaxies, which are red, highly concentrated,
and round; Late-type galaxies, which are blue, have low concentrations,
and are disk dominated; and Intermediate-type galaxies, which are red,
have intermediate concentrations, and have disks.
We have validated our method by comparing to visual
classifications of high-quality imaging data from the Millennium
Galaxy Catalogue.
The inclination correction is crucial to unveiling the
previously unrecognized Intermediate class.
Intermediate-type galaxies, roughly corresponding to lenticulars
and early spirals,
lie on the red sequence. The red sequence is therefore composed
of two distinct morphological types, 
suggesting that there are two distinct mechanisms
for transiting to the red sequence. We propose that Intermediate-type
galaxies are those that have lost their cold gas via strangulation,
while Early-type
galaxies are those that have experienced a major merger that either
consumed their cold gas, or whose merger progenitors were already
devoid of cold gas (the ``dry merger'' scenario).
\end{abstract}

\keywords{galaxies: fundamental parameters --- galaxies: structure}

\section{Introduction}
Galaxies do not smoothly occupy the full range of
possible structural and photometric parameter space. This fact was
first pointed out by \citet{hubble26}, whose ``Tuning Fork''
classified galaxies into ellipticals, spirals, and barred
spirals. The basic categorization of ``early''-type galaxies,
with a smooth quiescent elliptical appearance, and ``late''-type galaxies,
characterized by a spiral disk with sites of active star formation,
has remained to this day.

Many studies have attempted to put this classification on a
quantitative footing.
Bimodality in the color-magnitude plane has been well established
\citep[e.g.][]{strateva-etal01,blanton-etal03-properties,baldry-etal04,
balogh-etal04-colour}, and provides the most straightforward
classification method.
\citet{weinmann-etal06a} classified galaxies into three
types based on their colors and spectra: early-types, which
are red and passive; late-types, which are blue and active;
and intermediate-types, which are red and active.
These authors argue that their intermediate class are most
likely a combination of early-types whose star formation rates are
overestimated and reddened edge-on late-types.
\citet{ellis-etal05} statistically studied the multivariate distribution of
a large number of observed galaxy properties and found that exactly
two classes are required.
One particularly useful diagnostic plane for classifying galaxies
is color versus global concentration of the light profile,
where early-type galaxies appear red and concentrated while
late-type galaxies appear blue and have low concentrations
\citep[e.g.][]{driver-etal06}.

The use of such automated classifications is crucial in large galaxy
surveys such as those now available, for which visual examination 
of hundreds of thousands of galaxies is unfeasible%
\footnote{See, however, the Galaxy Zoo project:
http://galaxyzoo.org/.}. Moreover, automated methods are important
even when visual classifications are available, as they are
quantitative and allow finer and unbiased classification in cases
where the patterns identified by eye do not map correctly onto the
physically-important parameters.

Most observational parameters that are used in classification methods
are either calculated within a circular aperture or in circular
annuli. However, disk galaxies are not usually circular when viewed
on the sky, and their appearance varies systematically as a function
of inclination. For example, the front and back sides of a disk
contribute most of their light at small radii if the disk is highly
inclined, but at large radii if the disk is nearly face-on; inclined
disks therefore appear to be more concentrated.
Observed colors and magnitudes
are also affected by the increased dust absorption in highly inclined
disks relative to nearly face-on disks. Inclination effects therefore
introduce significant scatter in the observed parameters of individual
disk galaxies.

Fortunately, it is possible to calculate and correct for the effects
of inclination on several observed parameters.
In \citet{bailinharris08-trimodality} (Paper~I), a
simple inclination-independent measure of galaxy concentration was derived
based on the observed Petrosian concentration index \cpetro\ and
the observed isophotal axis ratio $b/a$
\citep[see also][who recalculated the concentration in elliptical
apertures directly from the galaxy images]{yamauchi-etal05},
Recently, four studies have used large samples of galaxies to
determine the effects of inclination on galaxy photometry:
\citet{shao-etal07} (hereafter \citetalias{shao-etal07}),
\citet{ur08} and \citet{maller-etal08},
all using the Sloan Digital Sky Survey (SDSS), and
\citet{driver-etal07}, using the Millennium Galaxy Catalogue (MGC),
measured how the characteristic turnover in the luminosity function
of spiral galaxies varies
as a function of inclination, and provide simple parametrizations
of the relative extinction in a galaxy as a function of disk inclination.
These new methods allow us to correct galaxy properties for
inclination effects and revisit the problem of automated galaxy
classification.

In this paper, we present a new method for classifying galaxies 
that uses
inclination-corrected values for galactic concentration, color,
and absolute magnitude, along with the apparent axis ratio.
We use the statistical power of SDSS, which
allows us to examine the multivariate distribution of galaxy properties
with ample signal-to-noise, and the MGC, which provides
deep high-resolution images for direct visual classification.
We describe the selection of the galaxy samples in section~\ref{sample-section}.
The expressions for the inclination-corrected
quantities are defined in section~\ref{face-on-values-section}\ and
a face-on color-magnitude diagram of SDSS galaxies is presented.
In section~\ref{classification-section} we present our method of
classifying galaxies and demonstrate its effectiveness, and finally
we present our conclusions in section~\ref{conclusions-section}.

\section{Galaxy Samples}\label{sample-section}
Our primary galaxy sample comes from SDSS,
a 5-band optical and near-infrared
imaging and spectroscopic survey covering one quarter of the sky
\citep{sdss-technical-summary}. The final data release,
DR6 \citep{sdss-dr6}, contains imaging and spectroscopy of over half
a million galaxies with well defined selection criteria. Our sample
consists of the $486305$ galaxies that meet the Main Galaxy
Sample targeting criteria \citep{sdss-maingalaxysample}, have
spectra that are classified as galaxies with confident redshifts
($z_{\mathrm{conf}}>0.85$), have dereddened Petrosian $r$-band
magnitudes $r<17.7$, lie in the redshift range $0.01 < z < 0.2$, and have
absolute magnitudes $\absmag < -17.5$. Petrosian magnitudes are
used throughout, and $r$-band quantities are used for all photometric
parameters. Colors and absolute magnitudes are k-corrected to $z=0$
with KCORRECT~v4\_1\_4 \citep{kcorrect}. We adopt
$\Omega_0=0.3$, $\Omega_{\Lambda}=0.7$,
and $H_0=70~\mathrm{km~s^{-1}~Mpc^{-1}}$.

While SDSS provides us with the statistics necessary to measure the
multivariate distribution of galactic parameters, the imaging is relatively
shallow and often taken in poor seeing conditions. Therefore, in order
to obtain confident visual classifications of a subset of galaxies,
we have used the deeper and higher resolution imaging
data from the MGC \citep{mgc-imaging}.
This $37.5~\mathrm{deg}^2$ $B$-band survey is entirely contained within
the SDSS footprint, allowing cross-identification of all galaxies.
We have randomly selected $400$ SDSS galaxies from the above sample
that are also contained in the cleaned bright MGC galaxy catalog%
\footnote{Note that the default matched mgc\_sdss catalog contains
objects that lie in MGC exclusion regions and are
therefore not suitable for scientific
analysis. We have obtained a version of the mgc\_sdss catalog
cleaned of such objects from J. Liske.}
for direct visual classification

\section{Correcting for Inclination}\label{face-on-values-section}
\subsection{Concentration}\label{face-on-concentration-section}
In Paper~I, we presented a method to correct the Petrosian concentration
\cpetro\ for inclination effects. We briefly review the method here.

As disk galaxies are seen closer to edge-on, the isophotal axis ratio
$b/a$ decreases while the global concentration
of the light profile increases. As a result, the loci of
intrinsically similar galaxies in
the \cpetro-$b/a$ plane curve to higher \cpetro\ at smaller
$b/a$. By constructing models of a pure exponential disk viewed at
a variety of inclinations, we have determined the expected
concentration of the disk $C_{\mathrm{Petro}}^{\mathrm{disk}}$
as a function of the observed isophotal axis ratio $b/a$.
We use this relationship to define the normalized
concentration, $\cnorm_{,i}$, for a galaxy with observed isophotal
axis ratio $(b/a)_i$ and observed Petrosian concentration $\cpetro_{,i}$ to be:
\begin{equation}
  \cnorm_{,i} \equiv \frac{\cpetro_{,i}}{C_{\mathrm{Petro}}^{\mathrm{disk}}
    (b/a_i)}.
\end{equation}
The loci of galaxies in the inclination-corrected \cnorm-$b/a$ plane no
longer depend on axis ratio, validating the use of \cnorm\ as
an inclination-independent measure of galaxy concentration.

Note that although elliptical galaxies do not suffer from the same
inclination effect as disks, they all have large apparent axis ratios.
At these large axis ratios, $C_{\mathrm{Petro}}^{\mathrm{disk}}$ is
almost independent of $b/a$, and therefore the magnitude of the
inclination correction is minimal. Therefore, applying the correction
to all galaxies does not introduce any bias in the concentrations
of elliptical galaxies.

\subsection{Magnitudes and colors}\label{face-on-magnitudes-section}
\citetalias{shao-etal07} find that the extinction in an inclined
disk galaxy relative to a face-on disk galaxy in a given band
is well described by:
\begin{equation}
\label{magcorrect-eq}
  M - M^{\mathrm{F}} = -\gamma_2 \log(\cos i)
\end{equation}
for observed absolute magnitude $M$, face-on
absolute magnitude $M^{\mathrm{F}}$, inclination $i$
(defined to be $0\degr$ for face-on disks and $90\degr$ for
edge-on disks), and
relative extinction coefficient $\gamma_2$. Values of
$\gamma_2$ for each SDSS band  are given in \citetalias{shao-etal07}.

In \citetalias{shao-etal07}, the inclination $i$ was determined from the
apparent axis ratio of the exponential component of the surface brightness
decomposition (``expab'' in the SDSS database)
by Monte Carlo simulations. We simply use the median mapping between
expab and $\cos i$ given in their figure~3. We then calculate the face-on
absolute $r$-band magnitude, \absmagfo, and
face-on \gr\ color, \grfo, using
equation~(\ref{magcorrect-eq}):
\begin{equation}
  \absmagfo = \absmag + \gamma_2^r \log(\cos i)
\end{equation}
\begin{equation}
  \grfo = \gr + (\gamma_2^g - \gamma_2^r) \log(\cos i).
\end{equation}

Correcting for the disk inclination
presumes the presence of a disk, while many galaxies are almost
entirely spheroidal systems.
Unlike the concentration
(\S~\ref{face-on-concentration-section}), the photometry is
sensitive to the inclination correction even at relatively large
axis ratios,
and should be omitted for elliptical galaxies.
We therefore only apply the correction to galaxies for which the bulge+disk
surface brightness decomposition in SDSS contains at least a $20\%$
disk, i.e. for which $\mathrm{fracdeV} < 0.8$. We include
galaxies where the bulge contributes a large fraction of the flux
(e.g. compared to \citetalias{shao-etal07}, who restrict their
analysis to $\mathrm{fracdeV} < 0.5$,
and \citealt{ur08}, who restrict their
analysis to $\mathrm{fracdeV} < 0.1$)
because the centrally-concentrated
bulge light is severely attenuated by disk dust even in the absence of
dust within the bulge itself, and
therefore the inclination correction is vitally important even in
systems with large bulges \citep{driver-etal07}.

\begin{figure*}
\plottwo{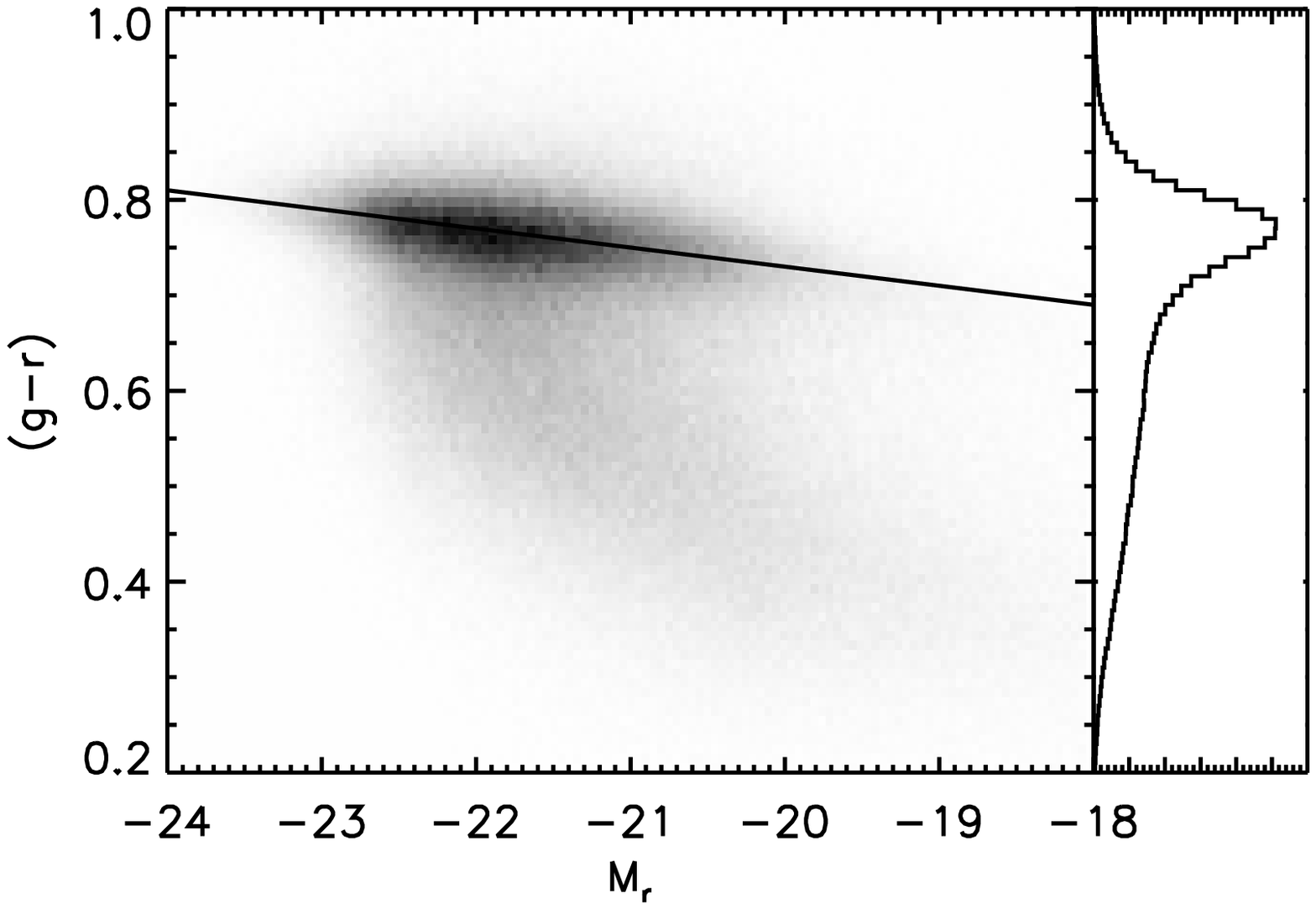}{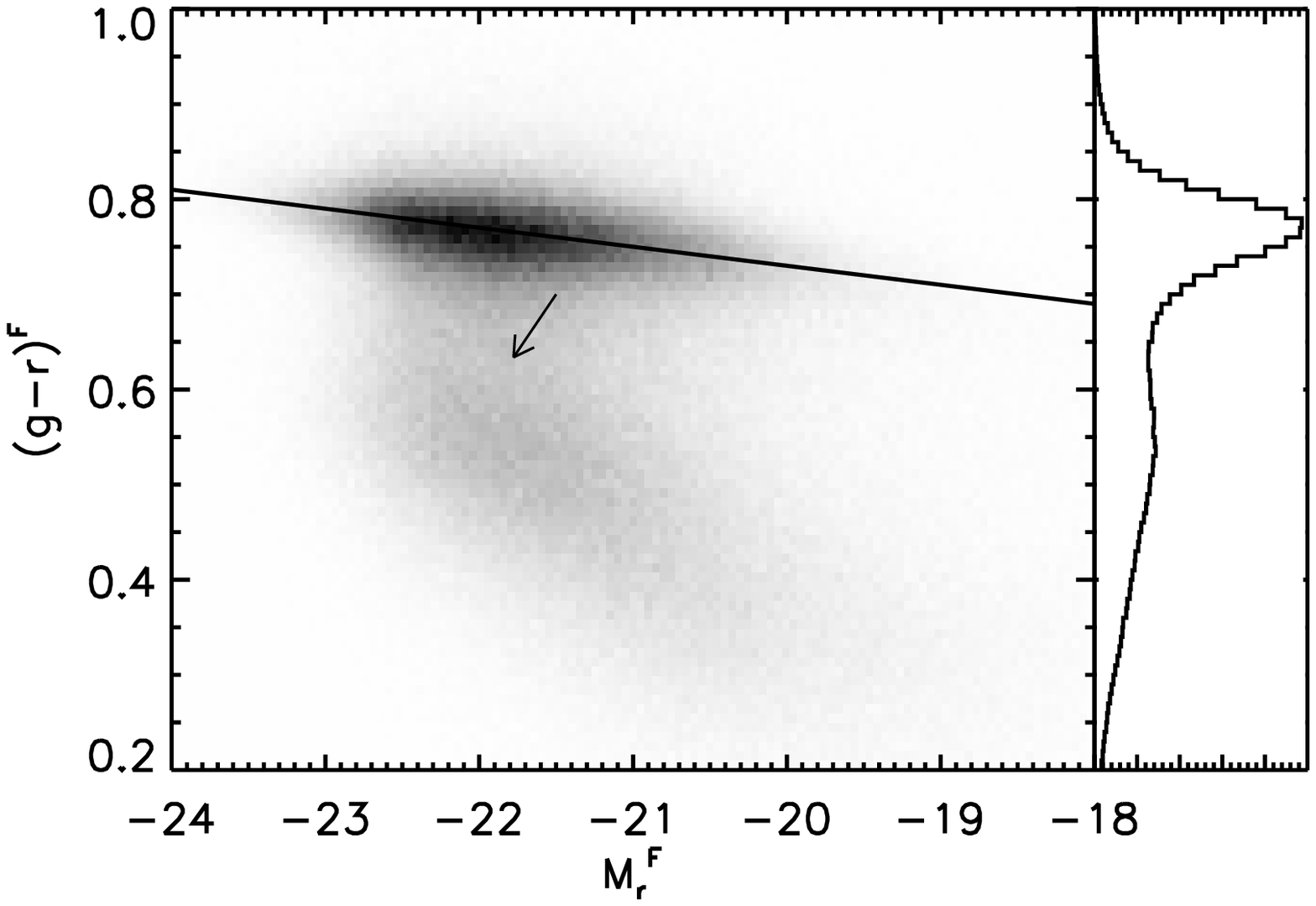}
\caption{\label{cmd}%
\textit{(Left)} Observed rest-frame \gr\ color versus \absmag\ color-magnitude
diagram for SDSS galaxies. The distribution of \gr\ marginalized over
absolute magnitude is shown in the histogram on the right. The solid
line denotes the red sequence.
\textit{(Right)} As in the left panel, but for face-on color \grfo\ versus
face-on absolute magnitude \absmagfo. The vector denotes the inclination
correction for a galaxy with $i=60\degr$.%
}
\end{figure*}

We plot the observed color-magnitude diagram (CMD) and the
inclination-corrected face-on color-magnitude diagram in Figure~\ref{cmd}.
The marginal distributions of \gr\ and \grfo\ are added as histograms.
The basic bimodality of the CMD is apparent in both plots, with a
red sequence and a blue cloud. In the observed CMD (left panel),
the blue cloud appears
to run into the red sequence at the luminous end and it is difficult to
distinguish red from blue galaxies in this region. In contrast, the
populations are much more clearly separated in the face-on CMD (right panel).
Similarly, while the marginalized distribution of \gr\ contains a red
peak with a blue tail, the marginalized distribution of \grfo\ contains
a distinct blue peak
(see also \citealp{maller-etal08}, who noted the increased separation
between red and blue galaxies; and \citealp{ur08}, who noted the
tightening of the blue sequence among galaxies with exponential profiles).

The effect of the inclination correction is directly demonstrated by
the vector in the right panel, which indicates the inclination correction
for a galaxy with $i=60\degr$. All inclination corrections occur parallel
to this vector.
The inclination correction moves galaxies almost perpendicular to the
red sequence, which is why it makes a relatively large difference in
the appearance of the CMD.

We have fit the red sequence by eye as
\begin{equation}
  (g-r)^{\mathrm{red}} = 0.77 + 0.02 (\absmag + 22).
\end{equation}
This is indicated by the solid line in both panels of Figure~\ref{cmd}.
We define the location of galaxy $k$ on the CMD relative to the red
sequence as its \cmd\ location parameter%
\footnote{Note that these definitions differ somewhat from the \cmd\ location
parameter used in Paper~I.}:
\begin{equation}
  \cmd_k = \gr_k - (g-r)^{\mathrm{red}}(\absmag_{,k})
\end{equation}
and analogously for the face-on CMD location parameter \cmdfo:
\begin{equation}
  \cmdfo_k = \grfo_k - (g-r)^{\mathrm{red}}(\absmagfo_{,k}).
\end{equation}

\section{Inclination-Independent Classification}\label{classification-section}
Traditionally, galaxies have been classified into two types:
``early-type'' or
``elliptical'' galaxies, which are red, round, smooth, and
have centrally-concentrated light profiles well approximated by
the \citet{devaucouleurs48} $R^{1/4}$ law or
the \citet{sersic-profile} profile with large values of $n$;
and ``late-type'' or ``spiral'' galaxies, which are blue,
disk-shaped, contain spiral and clumpy structure, and have
less concentrated exponential light profiles
or \citeauthor{sersic-profile} profiles with low values of $n$.
An intermediate
``lenticular'' or ``S0'' classification is also sometimes used, although
it is often considered an extension of either the elliptical or
spiral class depending on context.

In this section, we attempt to put these classifications on a
firm quantitative footing by using the inclination corrections
developed in section~\ref{face-on-values-section} to define an
inclination-independent method of classifying galaxies.

\subsection{Visual classification}\label{mgcclass-section}
We have randomly selected $400$ galaxies contained in both our
SDSS sample and the MGC and inspected the MGC images by eye. The advantages
of the MGC images are several: (1) they are deeper than the
SDSS images, allowing us to see fainter features in the outer regions
of the galaxy that can aid in classification, (2) the typical seeing
in the MGC is better than in SDSS, allowing effectively higher
resolution, and (3) by using single-band images we explicitly ensure
that our visual classification is purely based on the morphology
and has no color bias.

Each galaxy was given one of 5 classifications: ``elliptical'' (E),
``elliptical/lenticular'' (E/L), ``spiral/lenticular'' (S/L),
``spiral'' (S), or ``unclassifiable'' (X; usually a major merger).
Table~\ref{class-table} lists
the number of galaxies given each classification.

\begin{deluxetable}{lc|ccc}
\tablewidth{0pt}
\tablecaption{Comparison Between
Visual and Automated Classifications\label{class-table}}
\tablehead{%
 \colhead{Visual} & & \multicolumn{3}{c}{Automated Classification}\\
\colhead{Classification} & \colhead{Total} &
 \colhead{Early} & \colhead{Intermediate} & \colhead{Late}}
\startdata
E   & $133$ & $72$ & $49$ & $12$\\
E/L & $36$ & $15$ & $15$ & $6$\\
S/L & $45$ & $8$ & $17$ & $20$\\
S   & $174$ & $8$ & $13$ & $153$\\
X   & $12$ & $6$ & $3$ & $3$\\ \hline
Total & $400$ & $109$ & $97$ & $194$\\
\enddata
\end{deluxetable}

\subsection{Automated classification}\label{autoclass-section}

\begin{figure*}
\plottwo{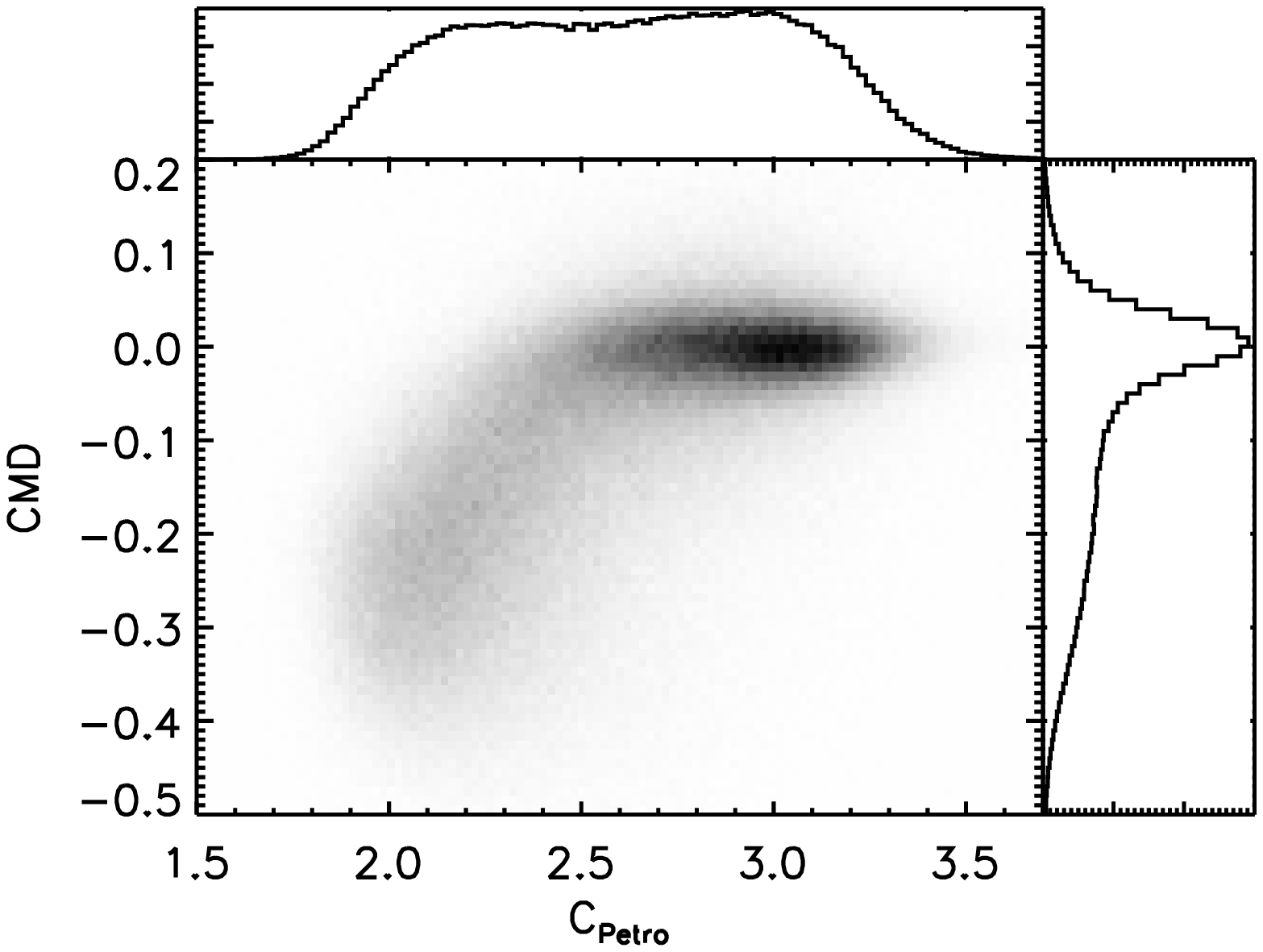}{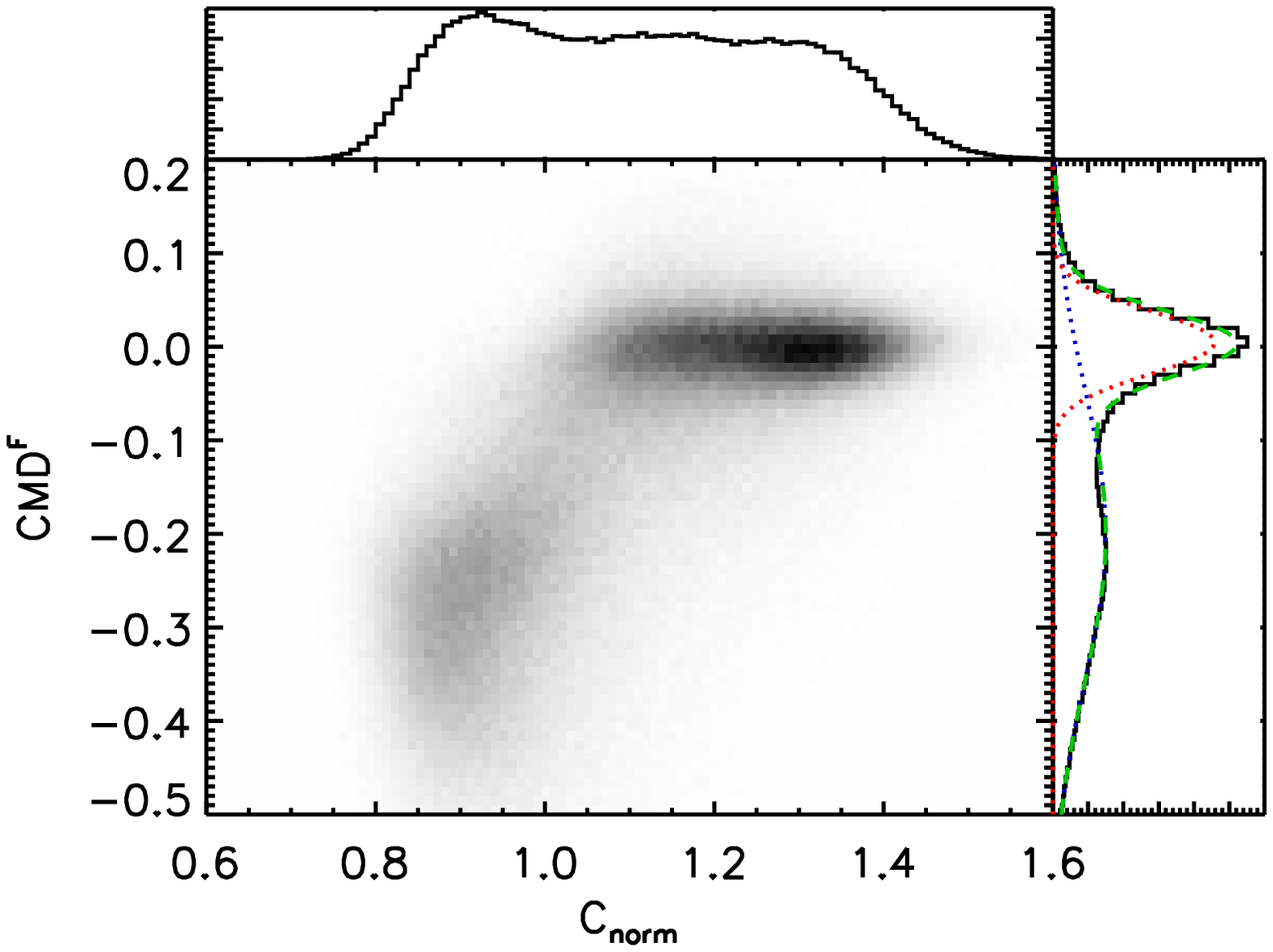}
%\plottwo{f2a.eps}{f2b_bw.eps}
\caption{\label{cmdcon}%
\textit{(Left)} Distribution of Petrosian concentration \cpetro\ versus
observed \cmd\ location parameter for SDSS galaxies. The marginalized
distributions are shown in the top histogram for \cpetro\ and the
right histogram for \cmd.
\textit{(Right)} As in the left panel for the inclination-corrected
concentration \cnorm\ and the face-on CMD location parameter \cmdfo.
The green dashed line in the \cmdfo\ histogram shows the bimodal Gaussian
fit to the distribution, while the red and blue dotted lines show the two
components of the fit.%
}
\end{figure*}

Our primary classification is performed in the color-concentration plane
\citep[e.g.][]{driver-etal06}. We adopt
the \cmd\ location parameter as the measure of color, as it provides
a better estimate of whether a galaxy lies on or off the red sequence
than \gr\ alone.
The joint distribution of observed concentrations \cpetro\ and observed
\cmd\ locations is shown in the left panel of Figure~\ref{cmdcon},
along with the distribution of each parameter independently. The tendency
for red galaxies to be more concentrated is apparent, as is an overall
bimodality of the galaxy population: there is a strong peak
of red galaxies at high concentrations ($2.5 \la \cpetro \la 3.4$)
and a more spread-out tail of blue galaxies at low concentrations.

The right panel of Figure~\ref{cmdcon} shows the joint and individual
distributions of the inclination-corrected concentration \cnorm\ and
the face-on color \cmdfo. The inclination correction makes several
appreciable improvements to this diagnostic plane. Firstly, as
described in Paper~I, the distribution of \cnorm\ is
\textit{tri}modal:
there is a distinct grouping of galaxies at intermediate
concentrations.
This trimodality has been previously unrecognized because inclination
effects smear together the intermediate- and high-concentration
subpopulations.
The color distributions, and therefore the luminosity-weighted stellar age,
of these two populations are similar, but they have distinct morphologies.
Secondly, the distribution of \cmdfo\ is much more bimodal
than that of \cmd: the previous blue tail that merged into the red sequence
has become a distinct peak.
We have employed the histogram fitting package RMIX
\citep[see also our more extensive description in Paper~I]{macd07}
to experiment with a range of possible multimodal fits that employ
relatively simple basis functions (Gaussian, lognormal, gamma
functions, etc.). Unlike the marginal distribution for \cnorm,
which clearly requires three components (see Paper~I), a bimodal
distribution fits the \cmdfo\ histogram successfully with no clear
evidence that would call for more than two components
\citep[see also][]{ellis-etal05}.
In Figure~\ref{cmdcon}b, we show the best-fit RMIX solution
that matches the \cmdfo\ distribution as the sum of two
Gaussian components. In this solution,
$60\%$ of the galaxies are assigned to the blue
peak while $40\%$ are assigned to the red peak.
Finally, the separation between classes in the joint distribution is much
stronger.
Rather than simply lying in
a spread-out tail, the late-type galaxies are relatively localized
in a distinct and well-separated region of parameter space from the
intermediate and early-type galaxies, which are themselves distinct.

\begin{figure*}
\plottwo{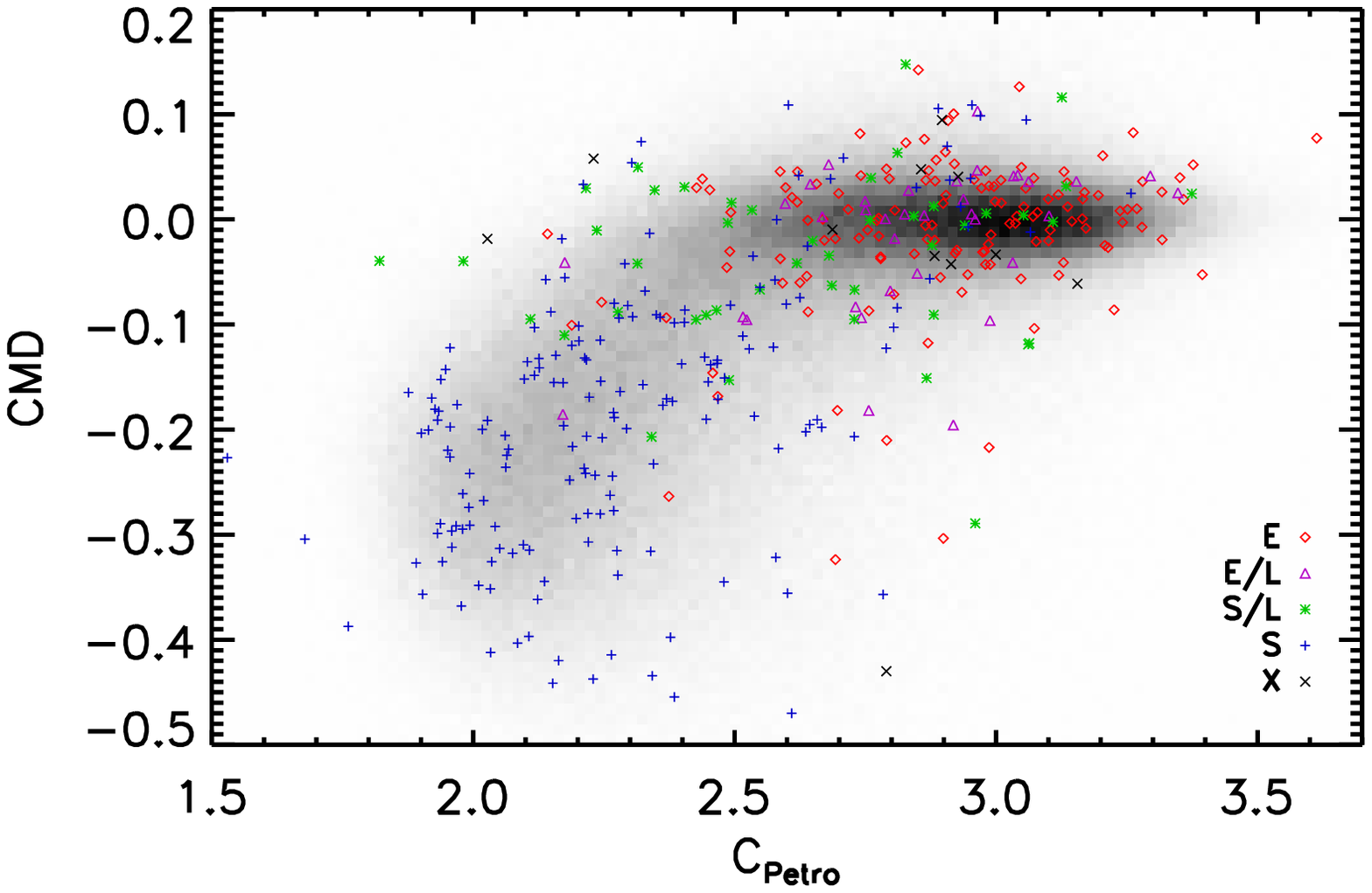}{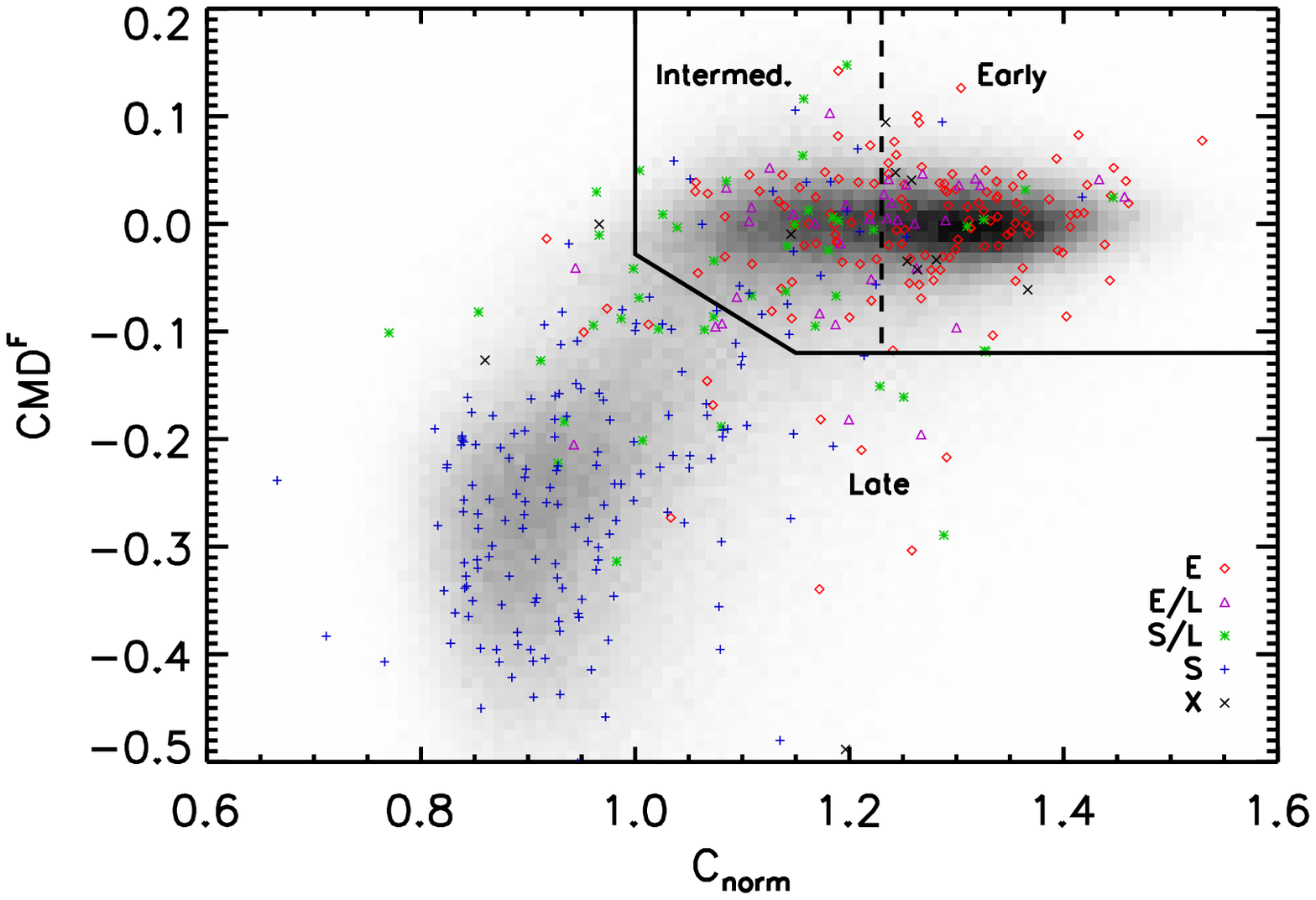}
%\plottwo{f3a_bw.eps}{f3b_bw.eps}
\caption{\label{cmdconeye}%
\textit{(Left)} The grayscale denotes the distribution of Petrosian
concentration parameters \cpetro\ versus observed \cmd\ location parameter
for SDSS galaxies, as in Figure~\ref{cmdcon}. The locations of the
visually classified galaxies are overplotted. The visual classifications
are: elliptical (``E'', red diamonds), elliptical/lenticular (``E/L'',
purple triangles), spiral/lenticular (``S/L'', green asterisks),
spiral (``S'', blue pluses), and unclassifiable (``X'', black crosses).
\textit{(Right)} As in the left panel for the inclination-corrected
concentration \cnorm\ and the face-on CMD location parameter \cmdfo.
The solid line denotes our adopted boundary between late-type galaxies
and intermediate/early-type galaxies, while the dashed line denotes
a possible boundary between intermediate and early-type galaxies based
on the trimodality of the \cnorm\ parameter.%
}
\end{figure*}

In Figure~\ref{cmdconeye} we have overplotted the locations of
the visually classified galaxies onto the \cmd-\cpetro\ and
\cmdfo-\cnorm\ distributions. The symbol types and colors represent the
visual classification. In both cases, the distribution at the
low-concentration blue end is dominated by spiral galaxies while the
high-concentration red end is dominated by elliptical galaxies.
In the uncorrected distribution, the intermediate type galaxies span
a wide range of parameter space, and in particular while elliptical
galaxies dominate the high \cpetro-high \cmd\ regime, there are also
a large number of lenticular galaxies and a non-negligible number of
spiral galaxies that also lie in this region of parameter space.
In contrast, the separation between the regions is much cleaner
in the inclination-corrected distribution. The inclination
correction moves galaxies to the left and downward, almost
completely evacuating the high \cnorm\ region of spirals and
spiral/lenticulars.

Based on the locations of visually classified galaxies on this plane
and on the shape of the joint distribution, we have drawn
a boundary in the \cnorm-\cmdfo\ plane to separate late-type galaxies
from intermediate- and early-type galaxies, indicated by the solid
line in the right panel of Figure~\ref{cmdconeye}. Within the
intermediate- and early-type region of parameter space, there is some
overlap in the distributions of the various visual classifications,
although there is a clear trend for more later types at smaller \cnorm.
We tentatively place a border between intermediate- and early-type
galaxies at $\cnorm=1.23$ (indicated by the dashed line), which is
the cross-over point between the intermediate and high-\cnorm\ peaks
in the \cnorm\ histogram (see Figure~\ref{cmdcon}).

\begin{figure}
\plotone{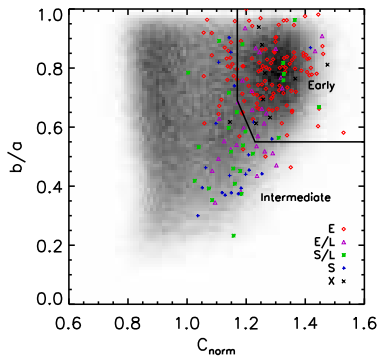}
%\plotone{f4_bw.eps}
\caption{\label{bacon}%
The grayscale denotes the distribution of inclination-corrected concentration
\cnorm\ versus isophotal $b/a$ axis ratio for SDSS galaxies
(see Paper~I). The locations of visually classified galaxies that lie
in the ``Intermediate'' and ``Early'' regions of Figure~\ref{cmdconeye}
are overplotted, with the same symbol types as in Figure~\ref{cmdconeye}.
The solid line denotes our adopted separation between intermediate-type
and early-type galaxies; this line is identical to the red line
defining the boundary of early-type galaxies in figure~2 of Paper~I.
}
\end{figure}

We have further examined the distribution of visually-classified galaxies
in the axis ratio-concentration ($b/a$-\cnorm) plane of Paper~I.
There are several reasons for doing this.
Firstly, the overlap in the distribution of visually-classified
galaxies of a given type within the Intermediate and Early regions
of Figure~\ref{cmdconeye} may perhaps be disentangled if the
populations differ in
another morphological parameter.
Secondly, we note that in Paper~I, we found that flattened galaxies with high
\cnorm\ have more in common with intermediate-concentration galaxies
than with ellipticals. While this may suggest a problem with our inclination
correction, the cause is simply statistics: the intrinsic axis ratio
distribution of ellipticals drops off dramatically at low $b/a$,
while the concentration distribution of intermediate-types is relatively
wide. Therefore, flattened galaxies with high \cnorm\ are more likely
to be unusually concentrated intermediate-types than unusually flattened
ellipticals.

In Figure~\ref{bacon}, we show the distribution of all SDSS galaxies
as the grayscale,
along with the visual classification of those galaxies
that lie in the ``Intermediate'' and ``Early'' regions of
Figure~\ref{cmdconeye}. The visual classification of these galaxies
varies systematically as a function of axis ratio even at
a constant concentration,
and the ellipticals are very concentrated in the high-\cnorm\ high-$b/a$
region of parameter space. We therefore use this plane to separate
early-type galaxies from intermediate-type galaxies as shown.

Our final classification is as follows:
\begin{itemize}
 \item Early-type galaxies lie to the upper-right of the solid
    boundary in Figure~\ref{cmdconeye} and to the upper-right of the
    boundary in Figure~\ref{bacon}.
 \item Intermediate-type galaxies lie to the upper-right of the solid
    boundary in Figure~\ref{cmdconeye} and to the bottom-left of the
    boundary in Figure~\ref{bacon}.
 \item Late-type galaxies lie to the bottom-left of the solid
    boundary in Figure~\ref{cmdconeye}.
\end{itemize}

A comparison of our automated and visual classifications for each
galaxy is given in Table~\ref{class-table}. It can be seen that
we do an excellent job of recovering the visual classification.
$80\%$ of our Early-type galaxies are classified by
eye as E or E/L, while $90\%$ of  Late-type galaxies are
classified as S or S/L by eye.
This is as good as could be expected given that
\citet{ellis-etal05} found that two people independently
visually classifying MGC images only agree at the $\sim 80\%$ level.

The Intermediate classification,
despite occupying a relatively limited region of parameter space,
contains galaxies that have a range of visual morphologies.
However, an examination of Figure~\ref{bacon} reveals that
the visual classification of these galaxies, which have intrinsically
very similar structural and photometric parameters, varies systematically
with apparent axis ratio, i.e. inclination. While
these galaxies contain disks, their passive appearance makes it
difficult to identify the disk structure when seen face-on.
Without the quantitative information available from the automated
classification, it would be impossible to distinguish these galaxies
from elliptical Early-type galaxies.

\begin{figure}
\plotone{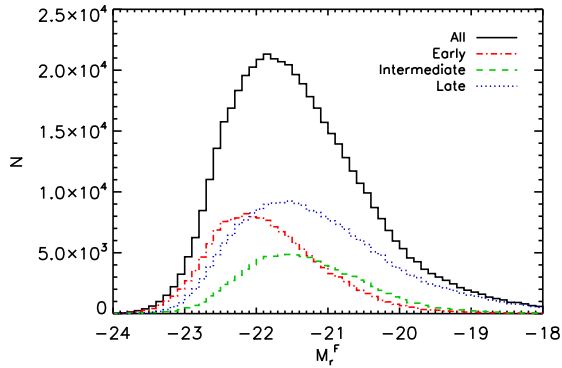}
%\plotone{f5_bw.eps}
\caption{\label{lumhist}%
Face-on absolute magnitude distribution of all SDSS galaxies (black/solid),
and those classified as Early (red/dot-dashed),
Intermediate (green/dashed), and Late (blue/dotted).}
\end{figure}

We evaluate the relative importance of each population
as a function of luminosity
by examining their absolute magnitude distribution
(Figure~\ref{lumhist}).
We recover the well-known result that the high-luminosity and low-luminosity
ends of the galaxy population are dominated by early- and late-type
galaxies respectively \citep[e.g.][]{nakamura-etal03}.
As suggested in Paper~I, the intermediate-type
galaxies have a narrower luminosity distribution.
Although a detailed analysis of the
luminosity functions is beyond the scope this paper, it is apparent
from Figure~\ref{lumhist} that the intermediate-type galaxies are
unimportant at the faintest magnitudes, and therefore the observed
intermediate types are representative of the population as a whole
and not simply the tip of a large low-luminosity ``iceberg''.

\section{Conclusions}\label{conclusions-section}
We have presented a new method to automatically classify galaxies
from the SDSS or any equivalent survey given four basic photometric
parameters: concentration, color, axis ratio, and absolute magnitude.
This method explicitly corrects for inclination
effects, allowing us to avoid the systematic misclassification
of inclined versus face-on systems endemic to most other automated
classifications.
We prefer this method even over visual classification, usually
considered to be the gold standard, which has considerable person-to-person
scatter \citep[see][]{ellis-etal05} and
is also susceptible to systematic inclination biases, as suggested by
the variation of visual classifications as a function of axis
ratio along the intermediate \cnorm\ locus in Figure~\ref{bacon}.

In Paper~I, we discovered that applying an inclination correction to
the observed concentration of SDSS galaxies
tightens an apparently bimodal distribution
into three distinct peaks. In this work, we have discovered that correcting
colors and absolute magnitudes for inclination effects also tightens the
``blue cloud'' of the CMD into a much more distinct peak in parameter space
\citep[see also][]{ur08}.
Based on the joint distribution of concentration and CMD
location, we find that the galaxy population separates well into
\textit{three} distinct classes, labeled here as ``Early'',
``Intermediate'', and ``Late''. We have validated our classification
scheme by comparing
to visual classifications of a subset of galaxies in the
high-quality MGC images.

A consequence of the results presented here and in Paper~I
is that there is an Intermediate galaxy type,
distinct from both Early-types (traditional ellipticals) and Late-types
(traditional spirals). These galaxies have intermediate concentrations
consistent with typical bulge-to-disk ratios of $\sim 3$,
a wide range of apparent axis ratios implying a disk morphology,
and colors indistinguishable from those of ellipticals of the same
luminosity (i.e. the \cmdfo\ distributions are similar, although this
corresponds to a bluer median color for the Intermediate types due to their
lower median luminosity).
Intermediate types span the full range of
visual classifications, although this may be more indicative of
the fallibility of visual classification than of the intrinsic properties
of Intermediate types.

It is interesting to contrast our galaxy types to those of
\citet{weinmann-etal06a}, who also identified an intermediate
class of galaxy. However, their intermediate types do not clump
in one particular region of parameter space but rather lie
at the intersection of the early- and late-type populations.
These authors consider that their intermediate-types may be
(a) edge-on reddened late-type galaxies, (b) early-type galaxies
whose star formation rates are overestimated, (c) a mixture of
(a) and (b) (their preferred explanation), or (d) a distinct population.
Our intermediate-types cannot be (a), because we have explicitly
corrected for inclination, nor (b) or (c), because the star formation
rate does not enter into our classification. Therefore, unlike
the intermediate-types identified by \citet{weinmann-etal06a},
our Intermediate classification must represent a truly distinct
galactic population. The lenticulars and bulgy disks that
constitute our Intermediate type are red and passive, and would
be classified as ``early'' in \citet{weinmann-etal06a}.

The standard explanation for the bimodality of the CMD is that
galaxies remain blue and star-forming while
they contain cold gas, but at some point in their evolution they
rapidly lose their cold gas
(perhaps due to ram pressure stripping, strangulation,
or galaxy harassment when entering a dense cluster or group environment;
removal of gas through a galactic wind or active galactic nucleus;
or exhaustion of their
fuel in a merger-induced starburst),
and quickly transit to the red sequence as the young stars fade.
Our discovery that there are two distinct populations on the red
sequence, an intermediate-concentration population that contains
a disk and a high-concentration population that is entirely spheroidal,
suggests that there are
two distinct mechanisms for transiting to the red sequence.
We propose that Intermediate-type galaxies are those whose gas-removal process
did not significantly perturb the global morphology of the galaxy
(e.g.~strangulation), while Early-type galaxies are those
whose gas was removed in a violent event like a major merger.
Alternatively, perhaps there is only one method of transiting
to the red sequence, which results in Intermediate-type galaxies,
while Early-types are formed from violent ``dry'' mergers of
Intermediate and Early-type galaxies.
Regardless, the presence of these distinct populations provides a
key observational clue to the processes that drive galaxy evolution.

It is important to note that our
samples consist of galaxies within the local universe,
i.e.~which have had a Hubble time for their structure to
regularize. Classification methods based on local galaxies might
not apply to younger high redshift galaxies.
Determining the epoch when these regular patterns emerged
would provide an important constraint in their origin.

\acknowledgements
JB thanks Joe Liske for help with the MGC database and the SDSS
help desk for their help with the Catalog Archive Server.

    Funding for the Sloan Digital Sky Survey (SDSS) and SDSS-II has been provided by the Alfred P. Sloan Foundation, the Participating Institutions, the National Science Foundation, the U.S. Department of Energy, the National Aeronautics and Space Administration, the Japanese Monbukagakusho, and the Max Planck Society, and the Higher Education Funding Council for England. The SDSS Web site is http://www.sdss.org/.

    The SDSS is managed by the Astrophysical Research Consortium (ARC) for the Participating Institutions. The Participating Institutions are the American Museum of Natural History, Astrophysical Institute Potsdam, University of Basel, University of Cambridge, Case Western Reserve University, The University of Chicago, Drexel University, Fermilab, the Institute for Advanced Study, the Japan Participation Group, The Johns Hopkins University, the Joint Institute for Nuclear Astrophysics, the Kavli Institute for Particle Astrophysics and Cosmology, the Korean Scientist Group, the Chinese Academy of Sciences (LAMOST), Los Alamos National Laboratory, the Max-Planck-Institute for Astronomy (MPIA), the Max-Planck-Institute for Astrophysics (MPA), New Mexico State University, Ohio State University, University of Pittsburgh, University of Portsmouth, Princeton University, the United States Naval Observatory, and the University of Washington.

The Millennium Galaxy Catalogue consists of imaging data from the
Isaac Newton Telescope and spectroscopic data from the Anglo
Australian Telescope, the ANU 2.3m, the ESO New Technology Telescope,
the Telescopio Nazionale Galileo and the Gemini North Telescope. The
survey has been supported through grants from the Particle Physics and
Astronomy Research Council (UK) and the Australian Research Council
(AUS). The data and data products are publicly available from
http://www.eso.org/$\sim$jliske/mgc/ or on request from J. Liske or
S.P. Driver.

{\it Facilities:} \facility{Sloan, ING:Newton, AAT, ATT, NTT, TNG, Gemini:Gillett}

\bibliography{/home/bailinj/papers/masterref.bib}

\end{document}